\documentclass[12pt,preprint]{aastex}

\newcommand{\iso}[2]{\hbox{${}^{#1}{\rm #2}$}}
\newcommand{\Msun}{\ensuremath{{M}_{\sun}}}
\newcommand{\Lsun}{\ensuremath{{L_{\sun}}}}

\newcommand{\appropto}{\mathrel{\vcenter{
  \offinterlineskip\halign{\hfil$##$\cr
    \propto\cr\noalign{\kern2pt}\sim\cr\noalign{\kern-2pt}}}}}

\shorttitle{Helium-enriched AGB models}
\shortauthors{Karakas et al.}

\begin{document}


\title{Nucleosynthesis in helium-enriched asymptotic giant branch models: Implications for
Heavy Element Enrichment in $\omega$ Centauri}


\author{Amanda I. Karakas, Anna F. Marino and David M. Nataf}
\affil{Research School of Astronomy and Astrophysics, 
Australian National University, Canberra, ACT 2611, Australia}
\email{Amanda.Karakas@anu.edu.au}




\begin{abstract}
We investigate the effect of helium enrichment on the evolution
and nucleosynthesis of low-mass asymptotic giant branch (AGB) stars
of 1.7$\Msun$ and 2.36$\Msun$ with a metallicity of $Z=0.0006$ 
([Fe/H] $\approx -1.4$). We calculate evolutionary sequences with 
the primordial helium abundance ($Y = 0.24$) and with helium-enriched 
compositions ($Y = 0.30, 0.35, 0.40$). For comparison we calculate
models of the same mass but at a lower metallicity $Z=0.0003$ 
([Fe/H] $\approx -1.8$) with $Y=0.24$.
Post-processing nucleosynthesis calculations are performed on each
of the evolutionary sequences to determine the production of elements 
from hydrogen through to bismuth. Elemental surface abundance predictions
and stellar yields are presented for each model. 
The models with enriched helium have shorter main sequence and AGB
lifetimes, and enter the AGB with a more massive hydrogen exhausted
core than the primordial helium model.  The main consequences are 1) low-mass
AGB models with enhanced helium will evolve more than twice as fast, 
giving them the chance to contribute sooner to the chemical evolution 
of the forming globular clusters, and 2) the stellar yields will be 
strongly reduced relative to their primordial helium counterparts. 
An increase of $\Delta Y = 0.10$ at a given mass decreases the yields of carbon by up 
to $\approx 60$\%, of fluorine by up to 80\%, and decreases the yields of the 
$s$-process elements barium and lanthanum by $\approx 45$\%. 
While the yields of first $s$-process peak elements strontium, 
yttrium and zirconium decrease by up to 50\%, the yields of rubidium either do 
not change or increase.
\end{abstract}





\section{Introduction}

The globular cluster (GC) $\omega$ Centauri (NGC 5139) shows a star-to-star spread in iron
that spans more than an order of magnitude, from $-2.2 \lesssim$[Fe/H]$\lesssim -0.7$ 
determined both by photometry and spectroscopy and with a mean metallicity of approximately 
[Fe/H] $\approx -1.7$ 
\citep{norris95,stanford06,kayser06,cjohnson08,calamida09,cjohnson10,simpson13}.
The spread in iron and other iron-peak elements is larger than measured in any
other Galactic GC including M22 \citep{dacosta11,marino11,marino12b}. 
Multiple populations have been observed along various evolutionary stages 
of the color-magnitude diagram (CMD) from the main sequence \citep{anderson97,bedin04,king12} through the subgiant
and red giant branches \citep[e.g.,][]{bellini10}, to the white dwarf cooling 
sequence \citep[e.g.,][]{calamida08,bellini13}.
Modeling of the CMD requires stellar isochrones with variations in iron of up to one dex and 
helium enhancements up to $Y \sim$ 0.40 \citep{norris04,piotto05,king12}.

Omega Cen also shows the abundance patterns between light elements (e.g., C, N, O, Na, Al) 
that are observed in mono-metallic GCs as well as variations in the Mg isotopes 
\citep{cjohnson09,cjohnson10,pancino11a,pancino11b,marino12a,simpson12,simpson13,dacosta13}.
The first evidence for helium enrichment came from photometry through the split main
sequence \citep{bedin04,norris04,piotto05}. \citet{dupree13} used the near-infrared transition 
of He I to obtain helium abundances for two red giant branch (RGB) stars, 
and find a difference of $\Delta Y \approx 0.17$, with the Al-rich star having the highest helium abundance.
These data and more illustrate that the chemical evolution of $\omega$ Cen shows a high 
level of complexity beyond that observed in any other GC.

Of particular importance in the understanding of the evolution of $\omega$ Cen is the relative ages 
between the various sub-populations. 
Numerous studies based on CMD analysis combined with the metallicity distribution have yielded conflicting results, with
an age spread between 0-5 Gyrs \citep{norris04,sollima05b,stanford06,villanova07}. The notable rise of heavy elements 
produced by the slow neutron capture process (the $s$-process, e.g., La, Ba) with increasing [Fe/H] 
\citep{norris95,cjohnson10,stanford10,marino11,dorazi11}  may require a star formation possibly extended over a few Gyrs 
\citep{smith00,marcolini07,romano07}.  On the other hand, isochrone fitting that takes into account helium
variations and C$+$N$+$O differences as observed in $\omega$ Cen \citep{marino12a} favors small age spreads 
between the different sub-populations \citep{herwig12,joo13}.
Rapid-formation scenarios wherein the entire cluster formed within a few times $10^{8}$ years have 
also recently been suggested by \citet{dantona11} and \citet{valcarce11} making this issue highly controversial.

The origin of the helium and light-element abundance correlations observed in $\omega$ Cen 
have been the subject of much debate. Both result from hot hydrogen burning \citep[e.g.,][]{prantzos07},
which produces helium via the CNO cycles as well as variations in C, N, O and F. Higher order hydrogen burning
reactions alter Na, Mg, and Al via the NeNa and MgAl chains. The two favored polluters include 
intermediate-mass asymptotic giant branch stars between $\approx 3-8\Msun$ \citep{ventura09a,dantona11}
and rapidly rotating massive stars \citep{decressin07,krause13}, although massive stars in
binaries have also been proposed \citep{demink09}. Numerous studies have discussed the merits and 
problems with each polluter and the various scenarios 
\citep[e.g.,][]{dantona02,fenner04,norris04,karakas06b,renzini08,decressin09,dercole10}.
The contribution of massive stars that explode as core collapse supernovae (Type II SNe) is required
owing to the spread in iron but the heavy elements have been attributed to the $s$-process acting in 
low and intermediate-mass AGB stars \citep{vanture94,smith00,cjohnson10,dorazi11}.  The mass range of AGB 
stars responsible for the neutron-capture elements in $\omega$ Cen is unknown but can be 
constrained from the variations in key elements affected by neutron density including 
Rb, Zr, Sr, Ba, and Pb. This conclusion is reached by considering how the two main neutron producing reactions
effect the predicted abundance distribution of heavy elements. The first reaction identified
was \iso{22}Ne($\alpha$,n)\iso{25}Mg, which occurs in massive stars and intermediate-mass AGB stars 
($M \gtrsim 3\Msun$) when temperatures exceed 300 $\times 10^{6}$K \citep{cameron60,truran77,cosner80}. 
The second is the \iso{13}C($\alpha$,n)\iso{16}O reaction that has been confirmed observationally 
and theoretically to be the main neutron source in low-mass AGB stars \citep[e.g.,][]{gallino98,abia01}. 
The \iso{22}Ne($\alpha$,n)\iso{25}Mg reaction is predicted to overproduce elements at 
the first $s$-process peak (e.g., Cu, Rb) over heavier elements such as Ba, La, and Pb. 
We provide a more detailed introduction to the evolution and nucleosynthesis of low-mass 
AGB stars in \S\ref{sec:lowmass}.

\citet{dorazi11} find that the Pb measurements in $\omega$ Cen stars suggest that the
peak stellar mass contributing to the production of heavy elements in $\omega$ Cen is not 
dominated by intermediate-mass AGB stars of $\approx 5\Msun$, which mostly produce Rb 
\citep{lugaro12,karakas12}, or massive stars which produce Cu, Rb and little Ba or Pb 
\citep[e.g.,][]{cunha02,pignatari10,frischknecht12}.
The Pb observations, extended star formation and possible younger ages of the metal-rich populations 
suggest that the peak mass contributing to the rise of the $s$ process in $\omega$ Cen is probably 
$M \lesssim 3\Msun$ \citep[see discussions in][]{stanford07}.

If low-mass AGB stars are contributing to the chemical evolution of heavy elements in the 
metal and {\em helium}-rich component of $\omega$ Cen, then those AGB stars will themselves be
metal and helium rich. The studies on the effect of helium enrichment on the 
evolution of low and intermediate-mass stars concentrate on the low-mass stars ($M \approx 0.8\Msun$) 
that influence the color-magnitude diagram of star clusters today 
\citep{sweigart87,gallart05,lee05,valcarce12,joo13,campbell13}.
Furthermore, most of the studies do not carry the evolution beyond the horizontal branch 
to the AGB where the richest nucleosynthesis occurs for stars less than 8$\Msun$. Here for the first
time we present helium-enriched stellar yields of stellar models from the main sequence through to the tip of the
AGB. We investigate the effect of helium enrichment on the stellar evolution through the giant branches
as well as the effect it has on the stellar yields of elements from hydrogen through to bismuth.
The focus of our study is on models of $Z = 0.0006$ (or [Fe/H] = $-1.4$) because the metal-rich
tail of $\omega$ Cen shows evidence for significant helium enrichment. We also calculate models at
the same mass at $Z=0.0003$ (or [Fe/H] = $-1.8$) with a primordial helium composition. At this metallicity
there is no evidence for a high helium enrichment and will be used for comparison to the metal-rich models.

In \S\ref{sec:lowmass} we begin with a description of low-mass AGB evolution and nucleosynthesis,
in \S\ref{sec:models} we describe the theoretical models and present the results of the stellar evolution
calculations, in \S\ref{sec:nucleo} we present the numerical method used to calculate the nucleosynthesis
predictions, and in \S\ref{sec:results} we present the stellar yields and nucleosynthesis results.
We finish with a discussion and concluding remarks in \S\ref{sec:discussion}.

\section{Low-mass Asymptotic Giant Branch Stars} \label{sec:lowmass}

Stars with masses less than about 8$\Msun$, depending on the global metallicity, $Z$, 
evolve through core hydrogen and helium burning before ascending the giant branch for the
second time. At this stage the star is said to be on the  asymptotic giant branch.
For the rest of this discussion and study we will focus on low-mass AGB stars, which are stars 
$\lesssim 3\Msun$ and do not experience hot bottom burning or the second dredge-up. 

The structure of an AGB star consists of an electron 
degenerate C-O core, surrounded by a He-burning shell, a He-rich intershell, and a H-burning shell 
\citep[for a review see][]{herwig05}.  The core and burning shells are surrounded by a large convective
envelope that is composed primarily of hydrogen. The He-burning shell thins
as the star evolves up the AGB and eventually becomes thermally unstable and flashes
or pulses every $10^{5}$ years or so, depending on the mass of the H-exhausted core
(hereafter core mass). The energy produced per thermal pulse (TP) is enormous 
(a few $10^{8}\Lsun$) and leads to vigorous 
convection in the He-intershell which mixes the products of He-nucleosynthesis 
throughout the entire intershell. The energy from the TP is converted into
mechanical energy and drives a strong expansion of the whole star. This pushes the H-shell
out to cooler regions where it is essentially extinguished.  The rapid cooling
caused by the expansion and the extinction of the H-shell may allow convection
to develop in regions previously mixed by the flash-driven convection in the He-intershell.
This inward movement of the convective envelope is known as third dredge-up (TDU), 
and is responsible for enriching the surface in \iso{12}C and other products of 
He-burning (e.g., F, \iso{22}Ne) as well as heavy elements produced by the $s$ process 
\citep[for a review of AGB nucleosynthesis see][]{busso99}. Following TDU, the star 
contracts and the H-shell is re-ignited, providing most of the surface luminosity for 
the next interpulse period. This sequence of TP--TDU mixing--interpulse phase
will occur many times on the AGB, depending on initial mass, metallicity, and the AGB 
mass-loss rate. 

The $s$ process occurs in the intershell of AGB stars where 
helium is abundant and the temperatures are high enough for ($\alpha$,n) reactions 
to occur.  There are two possible neutron sources in AGB stars: the 
\iso{13}C($\alpha$,n)\iso{16}O and the \iso{22}Ne($\alpha$,n)\iso{25}Mg reactions. 
The \iso{22}Ne neutron source is activated at temperatures higher than 
300$\times 10^{6}$\,K (MK). These high temperatures 
can only be reached inside the convective TPs of intermediate-mass AGB stars
\citep{truran77,lugaro12}.
The He-intershell of low-mass AGB models less than $\approx 2\Msun$ (at the metallicities
in this study) do not reach 300 MK, which means that neutrons are produced by 
the \iso{13}C($\alpha$,n)\iso{16}O reaction. The \iso{13}C neutron source is activated 
at lower temperatures \citep[e.g., 90 MK,][]{cameron55,straniero97} 
but requires an additional supply of \iso{13}C above that left from CN cycling to make it 
an efficient neutron source \citep[e.g.,][]{gallino98}.
For this reason some protons must be mixed into the top layers
of the He-intershell and this mostly likely occurs at the deepest
extent of TDU episodes, where a composition discontinuity forms between
the H-rich envelope and the He-rich intershell. The protons are readily 
captured by the abundant \iso{12}C to form \iso{13}C and \iso{14}N, 
resulting in the formation of a \iso{13}C pocket.

One of the largest unknowns in $s$ process calculations is the physical
mechanism that causes the formation of \iso{13}C pockets as well as the extent
in mass of the protons mixed into the intershell. This means that the size of the
\iso{13}C pocket can be essentially treated as a free parameter, although
the extent can be constrained using data from AGB stars, barium and CH stars, 
post-AGB stars, and planetary nebulae 
\citep[e.g.,][]{busso01,axel07a,shingles13}. We refer to \citet{lugaro12} 
for a detailed discussion \citep[see also][]{gallino98,goriely00,herwig05,cristallo09}.

\section{Helium-enriched Stellar Evolutionary Models} \label{sec:models}

In this study we evolve stellar models of mass 1.7 and 2.36$\Msun$ with global
metallicities of $Z= 0.0003$ and $Z=0.0006$ ([Fe/H] $\approx -1.8 \rm{\, and} -1.4$ 
respectively) from the pre-main sequence to the 
tip of the AGB. The mass of 1.7$\Msun$ was chosen as representative of a low-mass AGB
star that produces an $s$-process distribution typical of radiative \iso{13}C
burning \citep[see discussion in][]{lugaro12}. The 2.36$\Msun$ was chosen because
the core mass of the primordial model with $Y=0.24$ at the beginning of the AGB
is similar to the core mass of the helium-enriched 1.7$\Msun$ model with $Y=0.35$
and can therefore be used for comparison. Furthermore, all the 2.36$\Msun$ models 
show marginal activation of the \iso{22}Ne($\alpha$,n)\iso{25}Mg neutron source
and is therefore a borderline case between low and intermediate mass behavior.  

We calculate one model at each mass at $Z=0.0003$ with a primordial 
helium composition of $Y=0.24$. These lower metallicity models are evolved 
for comparison to the models of $Z=0.0006$, which are the focus of this study. 
At $Z=0.0006$ we evolve four separate stellar models with initial helium abundances
of $Y = 0.24, 0.30, 0.35$ and $Y = 0.40$. The $Y = 0.24$ model will be referred
to as the primordial helium model. While varying helium we keep the global
metallicity constant at $Z=0.0006$ which means that the initial hydrogen
abundance decreases. The initial composition of C, N, and O are scaled solar
and the same in all the $Z=0.0006$ model calculations (note that we 
consider an $\alpha$-enhancement of $+0.4$~dex on the nucleosynthesis of 
the primordial models). The stellar models are 
calculated using an updated version of the Mount Stromlo Stellar Evolution Code 
\citep{lattanzio86,frost96,karakas07b,karakas10b}. We use the same version of 
the code described in \citet{kamath12} which 
includes the latest C, N-rich low temperature opacity tables from AESOPUS \citep{marigo09}.
In contrast to the models of \citet{kamath12} no convective overshoot
was applied to the border of the convective envelope in order to obtain third dredge-up.
Convection is approximated using the Mixing-length Theory with a mixing-length
parameter of $\alpha = 1.86$ in all calculations. We assume no mass loss on the
RGB and use the \citet{vw93} mass-loss formulation on the AGB. 

\begin{figure}
\begin{center}
\includegraphics[width=10cm,angle=270]{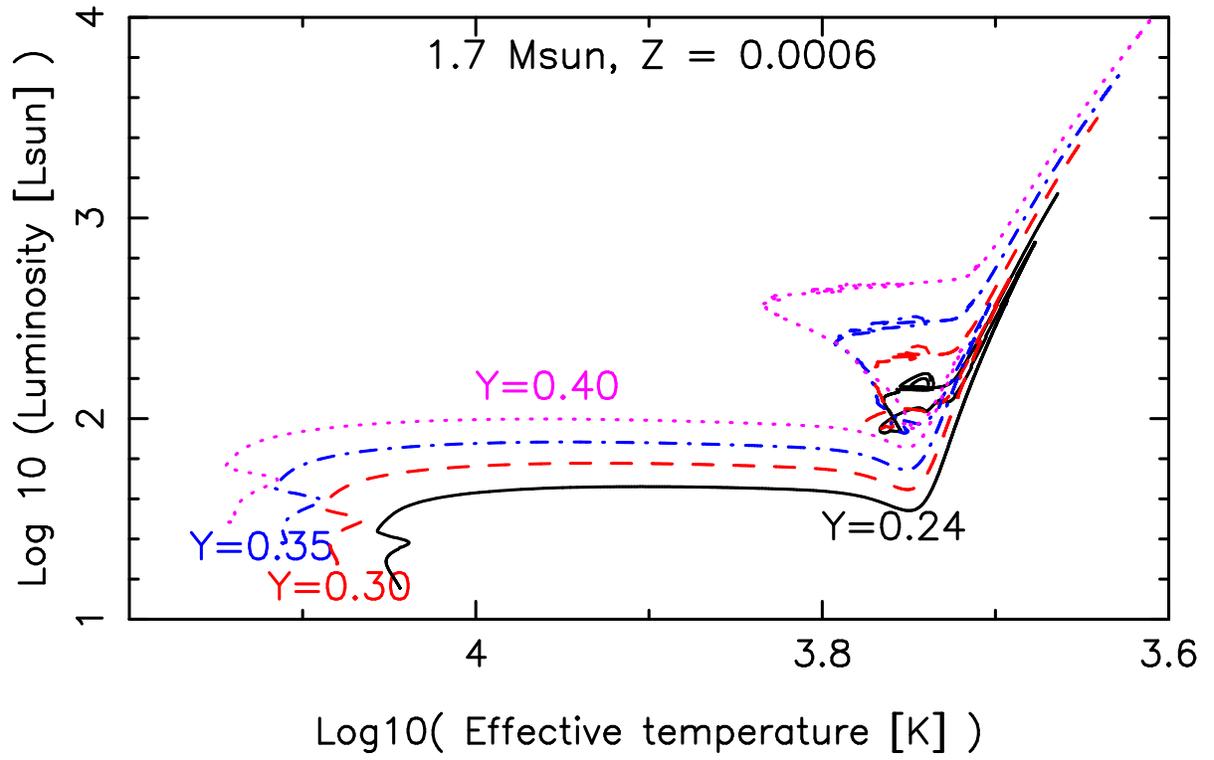} 
\caption{Evolutionary tracks of the four stellar models
of 1.7$\Msun$, $Z = 0.0006$. The initial helium composition of each track
is noted on the figure.}
\label{fig1}
\end{center}
\end{figure}

In Figure~\ref{fig1} we show the evolutionary tracks of the 1.7$\Msun$, $Z=0.0006$ 
models from the main sequence to the start of the thermally-pulsing AGB. 
This figure illustrates the effect
of an enhanced helium abundance on the surface luminosity and effective temperature
as a function of evolution. At a given evolutionary phase the model with increased helium
is hotter and more luminous. The hotter temperatures and higher luminosities arise because 
of the higher mean molecular weight imposed by the enhanced helium mass fraction 
at the beginning of the main sequence compared to the primordial helium model.
Higher temperatures also lead to a larger hydrogen exhausted core at the end
of the main sequence and this carries through to the beginning of the AGB.  

Table~\ref{table1} is a summary of the properties of the stellar models.
We include the main sequence lifetime, $\tau_{\rm ms}$; the RGB lifetime, $\tau_{\rm rgb}$
and the core helium burning lifetime, $\tau_{\rm coreHe}$. We then show 
the core mass at the first TP, $M_{\rm c}^{\rm 1st}$; the core mass at the
first TDU episode, $M_{\rm c}^{\rm TDU}$; the total number of thermal pulses
during the AGB; the total mass dredged up into the envelope, $M_{\rm dredge}^{\rm tot}$;
the maximum dredge-up efficiency, $\lambda_{\rm max}$; which is calculated according
to $\lambda = \Delta M_{\rm dredge}/\Delta M_{\rm c}$, where $\Delta M_{\rm dredge}$
is the amount of material dredged into the envelope and
$\Delta M_{\rm c}$ is the amount by which the H-exhausted core grew during the
preceding interpulse phase. We then include the maximum temperature reached during
a TP, $T_{\rm TP}$; and the maximum mass of the flash-driven convective 
zone in the He-intershell at the last TP, $M_{\rm Heshell}^{\rm f}$, which is a good 
approximation to the total mass of the He-intershell at the final TP.  
The final stellar mass, core mass, and envelope mass ($M_{\rm env}$)
at the last time step are included, along with the total stellar lifetime, $\tau_{\rm stellar}$.
For the 2.36$\Msun$ models we also include the maximum temperature at the base 
of the convective envelope, $T_{\rm bce}$.  Lifetimes are in Myr ($10^{6}$ years),
masses in solar units ($\Msun$) and temperatures in Kelvin.

For the 1.7$\Msun$ models, the envelope masses for each stellar evolutionary 
sequence are small ($M_{\rm env} \lesssim 0.1\Msun$) except for the $Y=0.40$ model.
The $Y=0.40$ model has $\approx 0.2\Msun$ of envelope left, which is less than what
was lost between the last two TPs ($\Delta M \approx 0.35\Msun$). 
The primordial models are evolved to the white dwarf cooling track and have 
no envelope left.  A small final $M_{\rm env}$ means that all mixing episodes
that will affect the final expelled yields have been calculated. This is the
case of the 1.7$\Msun$ models, and the primordial 2.36$\Msun$ models which
have $\approx 0.1\Msun$ of envelope left. Calculations of the helium-enhanced 
2.36$\Msun$ models experienced convergence problems before the models lost their
envelopes. These models lose $\Delta M \approx 0.3\Msun$ of envelope between TPs
during the superwind, which indicates that the $Y=0.30, 0.35$ models may experience
one more TP, while the $Y=0.40$ may experience up to two more TPs. The occurrence
of thermal pulses do not always lead to TDU and there is some evidence that 
the TDU efficiency, $\lambda$, decreases with decreasing $M_{\rm env}$
\citep[e.g.,][]{straniero97,karakas02}. Given the uncertainty it is unclear
how many more mixing episodes would occur. For these models we take the stellar
yields as lower limits. Note that the stellar lifetimes in Table~\ref{table1}
are accurate even if we miss one or two TPs. This is because the interpulse
periods are short relative to the total stellar lifetime and missing two means
we are missing $\approx 0.04$~Myr for the 1.7$\Msun$ models and $\lesssim 0.02$~Myr
for the 2.36$\Msun$ models.

The models with higher helium appear to evolve as more massive
stars of the same metallicity during their pre-AGB evolution and this is also
somewhat the case during the AGB. The models with increasing
helium show thinner helium intershells (as measured in mass), higher helium-shell 
burning temperatures, and even mild hot bottom burning in the case of the 2.36$\Msun$ model 
with $Y= 0.40$. This model also experiences second dredge-up after core 
helium burning, the only model to do so.  For the models of 1.7$\Msun$ we 
also find increasing interior  temperatures and core masses in models of 
increasing helium. While the dredge-up efficiencies and helium-shell burning
temperatures increase in all helium-enhanced models, they do not increase as much as
we would expect in a star of the same core mass but born with a primordial
composition (and therefore higher total mass). To illustrate this, the 1.7$\Msun$ model 
with $Y= 0.35$ enters the TP-AGB with a similar core mass to the 2.36$\Msun$, $Y=0.24$ model.
Helium shell temperatures in the 1.7$\Msun$ model with $Y=0.35$ never exceed 300~MK and 
$\lambda_{\rm max} \approx 0.6$.   In contrast, by the 8$^{\rm th}$ TP, the 2.36$\Msun$ model 
with $Y=0.24$ has peak temperatures of 318~MK and dredge-up efficiencies 
$\lambda \ge 0.9$.

The amount of material mixed into the envelope during the AGB is a crucial
parameter for determining the level of chemical enrichment. 
Interestingly models with the same mass, helium content but different $Z$ 
(comparing the $Z=0.0006$ to 0.0003 models, so a factor of two change),
show a similar AGB evolution, dredging up approximately the same amount of material
and finishing with roughly the same core mass. Helium on the other hand,
is a much more significant parameter. From Table~\ref{table1} we see
that the total amount of matter dredged into the envelope is lower in the
helium-enriched models. For example, the 1.7$\Msun$ model with $Y=0.40$ dredges
up a factor of $\approx 4$ less helium-shell material during the AGB while the
2.36$\Msun$ model with $Y=0.40$ dredges almost a factor of 6 less material (while
experiencing more thermal pulses).  The helium-enhanced 1.7$\Msun$ models 
experience fewer TPs than the primordial model; this combined
with smaller He-intershells reduces the total amount of dredge-up material.
The reduction in the mass of the He-intershell is particularly significant
in the helium-enhanced 2.36$\Msun$ models where the $Y=0.40$ model has a
He-intershell that is a factor of $\approx 4$ lower in mass than the primordial
model (c.f. only a factor of 2 in the 1.7$\Msun$). This reduction in
He-shell mass is the main reason for the reduction in total dredged-up material,
because as shown by Table~\ref{table1} the dredge-up efficiency stays approximately
the same at $\lambda \approx 0.9$ in the 2.36$\Msun$ models.

Also crucial for the chemical enrichment is the total stellar lifetime.
From Table~\ref{table1} we see that the lifetime of the helium-enriched
1.7$\Msun$ models are less than 1~Gyr, whereas the lifetime of the 
2.36$\Msun$ helium-enriched models ($\lesssim 300$~Myr) become comparable to 
intermediate-mass AGB stars of $\gtrsim 3\Msun$. This indicates that while 
low-mass primordial composition AGB stars with $M \lesssim 2\Msun$ may not have time to 
contribute to the chemical evolution of $\omega$ Cen, those stars born out
of helium and metal-enriched mixtures will. These stars will also contribute
less material to the interstellar medium. This could lead to a plateau
in the chemical evolution of $s$-process elements, as found by 
\citet{norris95} as we discuss in more detail in \S\ref{sec:discussion}.

\section{Stellar Nucleosynthesis Models} \label{sec:nucleo}

To study the nucleosynthesis of the elements up to bismuth we use a
post-processing code that takes the stellar evolutionary sequences as input.
The stellar inputs are the temperature, density, and convective velocities (for convective regions) 
at each mass shell in the model star as a function of time. Convective velocities 
are needed because the code includes in the equations to be solved for the abundances both the 
changes due to nuclear reactions and those due to mixing. The code has been previously described 
in detail by, e.g., \citet{cannon93}, \citet{lugaro12}, and \citet{karakas12}. The nuclear network
we use is a network of 320 species and 2,336 reactions, where most of the isotopes are located along the 
valley of $\beta$-stability.  The nuclear reaction network is based
on the JINA REACLIB database as of 2012 \citep{cyburt10}.

The $Z=0.0006$ models have an [Fe/H] $\approx -1.4$ whereas the $Z = 0.0003$ models have 
[Fe/H] $\approx -1.8$ using $\log \epsilon$(Fe)$_{\odot}$ = 7.54 and the standard spectroscopic 
notation where [X/Fe] = $\log_{10}$(X/Fe)$_{\rm \*} -\log_{10}$(X/Fe)$_{\odot}$.
This places the $Z =0.0006$ models near the metal-rich tail of stars in $\omega$ Cen 
\citep[e.g.,][]{cjohnson08}. Note that we use the proto-solar abundances for C, N, O, and Fe 
from Table~5 of \citet{asplund09}. In the post-processing calculations we take the 
initial hydrogen and helium abundances from the evolution calculations.  For the 
primordial helium models, we set the initial abundances to be scaled solar, with
the exception of the $\alpha$-elements [O, Ne, Mg, Si/Fe] = $+0.4$. For the
helium-enriched models we set all initial abundances to be scaled solar. 
We base these choices of the initial C, N, and O on the observed abundances of $\omega$ 
Cen stars from \citet{marino12a}.
The abundances of elements lighter than Fe do not affect the predictions of 
$s$-process elements but can provide some indication of the level of pollution expected. 

The He-intershell of the 1.7$\Msun$ models do not reach 300 MK (see Table~\ref{table1})
which means that neutrons only come from the \iso{13}C($\alpha$,n)\iso{16}O reaction.
For the 2.36$\Msun$ model the temperature in the He-shell reaches over 300~MK which
also leads to activation of the \iso{22}Ne($\alpha$,n)\iso{25}Mg neutron source.
This is measurable by the change in isotopic ratios of Mg as well as the ratios of Rb to
Sr and Zr \citep{abia02,karakas12}.
The inclusion of the \iso{13}C neutron source is performed during the post-processing by 
forcing the code to mix a small amount of protons from the envelope into the intershell 
at the deepest extent of each TDU. The procedure we use is the same as outlined in detail by 
\citet{lugaro12} \citep[and similar to][]{goriely00} where we apply the assumption that the proton 
abundance in the intershell 
decreases from the envelope value of $\simeq$ 0.7 to a minimum value of 10$^{-4}$  at a 
given point in mass located at ``$M_{\rm mix}$'' below the base of the envelope. We keep the value
of $M_{\rm mix}$ constant from pulse to pulse.  The value 
of $M_{\rm mix}$ is chosen such that the resultant \iso{13}C pocket is about 1/10$^{\rm th}$ of the
mass of the He-intershell, and is set at $M_{\rm mix} = 0.001, 0.002, 0.004\Msun$ in the 1.7$\Msun$
models and $M_{\rm mix} = 1\times 10^{-4}, 0.001, 0.002\Msun$ in the 2.36$\Msun$ models.
In the most helium-rich 1.7$\Msun$ and 2.36$\Msun$ models with $M_{\rm mix}$ = 0.004$\Msun$ 
and 0.002$\Msun$ respectively, the resultant \iso{13}C pocket may extend to cover up to 20\% 
of the helium intershell, owing to a thinner shell relative to the primordial model.

While we treat the size of the partially mixed zone as a free parameter, it can be constrained by
comparison to observations. Observations suggest that stochastic variations in the effectiveness
(or size) of the \iso{13}C pocket in AGB stars are present. \citet{axel07a} find that galactic disk 
objects are reproduced by a spread of a factor of 2--3 in the effectiveness of the \iso{13}C pocket, 
whereas \citet{busso01} required a spread of a factor of $\approx 20$.  
Comparisons to lower metallicity post-AGB stars require larger spreads of a factor of 3--6 
\citep{axel07b,desmedt12}. 
Comparisons of theoretical predictions to observed abundances of carbon enhanced metal-poor stars 
suggest that the
size of the \iso{13}C pocket could vary up to a factor of 10 or more \citep{bisterzo11,lugaro12}.
In summary, there is observational evidence that a spread in effectiveness of the \iso{13}C pocket
is needed in theoretical models, but there is no consensus on how large that spread actually is. This
problem indicates a significant lack of understanding on the mechanism(s) responsible for the 
formation of \iso{13}C pockets in AGB stars and is related to the treatment of 
convection in stellar codes \citep[see discussions in][]{goriely00,lattanzio05,herwig05}. 
Note that we do not consider the effect of rotation on AGB models, which are known to 
affect $s$ process predictions \citep{herwig03,piersanti13}.

\section{Nucleosynthesis and Stellar Yields} \label{sec:results}

We calculate the elemental stellar yields, which are available as on-line data tables for all 
stellar evolutionary sequences listed in Table~\ref{table1}.  We provide two tables for download, 
one containing all the yields for the $Z = 0.0003$ models (called ``z0003.yields.tab'') and another 
for all the $Z = 0.0006$ models in this study (called ``z0006.yields.tab'').
All the data, including additional data tables for $Z = 0.0003$ models, are available in a .tar.gz 
package in the electronic edition.
Table~\ref{table2} shows a portion of the yield table for the $Z=0.0006$ models and is published in its 
entirety in the electronic edition. Each table begins with a table header that lists the initial mass 
(in $\Msun$), metallicity, helium content, and size of $M_{\rm mix}$ used in the calculations. 
The second row is another header that provides the final mass (in $\Msun$) and the total mass expelled 
into the interstellar medium (ISM) (in $\Msun$).  The total expelled mass is simply the initial mass 
minus the final core mass,  using values from Table~\ref{table1}.
The columns in Table~\ref{table2} are the element symbol, the atomic number ($Z$), and the 
abundance in terms of $\log \epsilon(X) = \log_{10}(X/H) + 12$, 
where $X$ is the abundance by number of element $X$. Next we provide the [X/H] and [X/Fe] ratios 
and the mass fraction of element X. All abundances are calculated from the 
average composition of the ejected stellar wind. 

In the final column of Table~\ref{table2} we include the stellar yield, which we define 
here as the mass of X expelled into the ISM over the stellar lifetime.  
The mass expelled into the ISM is calculated according to
\begin{equation}
 M_{\rm yield} = \int_{0}^{\tau} \left[ X(t) \right] \frac{d M}{dt} dt,
\label{yield}
\end{equation}
where $M_{\rm yield}$ is the yield of species $X$ (in solar masses), $dM/dt$ is the current 
mass-loss rate, $X(t)$ is the current mass fraction of species $X$,  and $\tau$ is the total lifetime 
of the stellar model. The yield as defined here is always positive. 

We include elemental yields for all stable elements except Li, Be, B; the radioactive Tc is included because it 
is a tracer of the $s$ process but the un-stable elements Pm and Po are not included. We have assumed that 
radioactive species have decayed to their daughter products (e.g., \iso{26}Al has decayed to \iso{26}Mg) with
the exception of Tc.  After the elemental yields are listed, we also 
include the $s$-process indicators, which are obtained from the ejected composition of the stellar wind: 
[Rb/Zr], [ls/Fe], [hs/Fe], [hs/ls], and [Pb/hs], where [ls/Fe] and [hs/Fe] are defined according to 
[ls/Fe] = ([Sr/Fe] + [Y/Fe] + [Zr/Fe)/3, and [hs/Fe] = ([Ba/Fe] + [La/Fe] + [Ce/Fe])/3.

\subsection{Summary of Abundance Predictions}
 
In Table~\ref{table3} we show representative predicted abundance ratios (in [X/Fe]) of elements 
produced by AGB stars including C, N, F, and Na as well as the neutron-capture elements 
Kr, Rb, and Pb, and the $s$ process indicators [ls/Fe] and [hs/Fe].  Abundance predictions are
from the $Z=0.0006$ models with different helium abundances, as noted in the table, and 
calculated with the same $M_{\rm mix}= 0.001\Msun$.

Table~\ref{table3} shows that
the abundances of the light elements C, N, F, and Na are enhanced by AGB nucleosynthesis in all 
calculations to some extent. The elements traditionally synthesized by AGB models such as C and F are 
particularly enriched (at the level of [X/Fe] $\gtrsim 1.5$). Other elements (e.g., N, O, Na) are 
only marginally enhanced at the level of $\lesssim 0.5$~dex (e.g., oxygen). Mg and Al are 
not produced in the 1.7$\Msun$ models and marginally produced at 2.36$\Msun$ at the level of 
$\lesssim 0.4$ dex for Mg and 0.15~dex for Al. The intermediate-mass elements from Si to Fe are not
produced by AGB stars \citep{karakas09,cristallo11} although P and Sc can be made by neutron captures 
during convective thermal pulses. We find maximum enhancements on the level of 0.55 dex for 
P and about 0.30~dex for Sc. Iron group elements are not produced by AGB stars and although some Fe
is consumed by the $s$ process, the overall change to the elemental Fe abundance is very small. The 
enhancement of elements heavier the Ni, including Cu and Zn which are at the beginning of the $s$-process 
path, show some increase depending on the mass, helium content, and size of $M_{\rm mix}$. 
Eu and other elements associated with the $r$ process (e.g., Pt, Au) are enhanced on the level of
$\lesssim 1$ dex. 
Although the chemical evolution of these elements is dominated by the $r$ process, low-metallicity
low-mass AGB stars can synthesize some amount of e.g., Eu \citep{cristallo09,bisterzo10,lugaro12}. 
The amounts produced are always lower than traditional $s$-process elements (Sr, Ba, Pb), e.g., 
[Ba/Eu] = 1.0 and [Pb/Eu] $\approx 1.8$ in the 1.7$\Msun$ models. 

The two lower metallicity 1.7$\Msun$ and 2.36$\Msun$ models with $Y=0.24$ and $Z=0.0003$ produce
a similar nucleosynthesis abundance pattern to their more metal-rich counterparts at $Y=0.24$.
In all cases we find a stronger level of chemical enrichment in the lower metallicity models, 
when measured by the [X/Fe] ratios.

\begin{figure}
\includegraphics[width=10cm,angle=270]{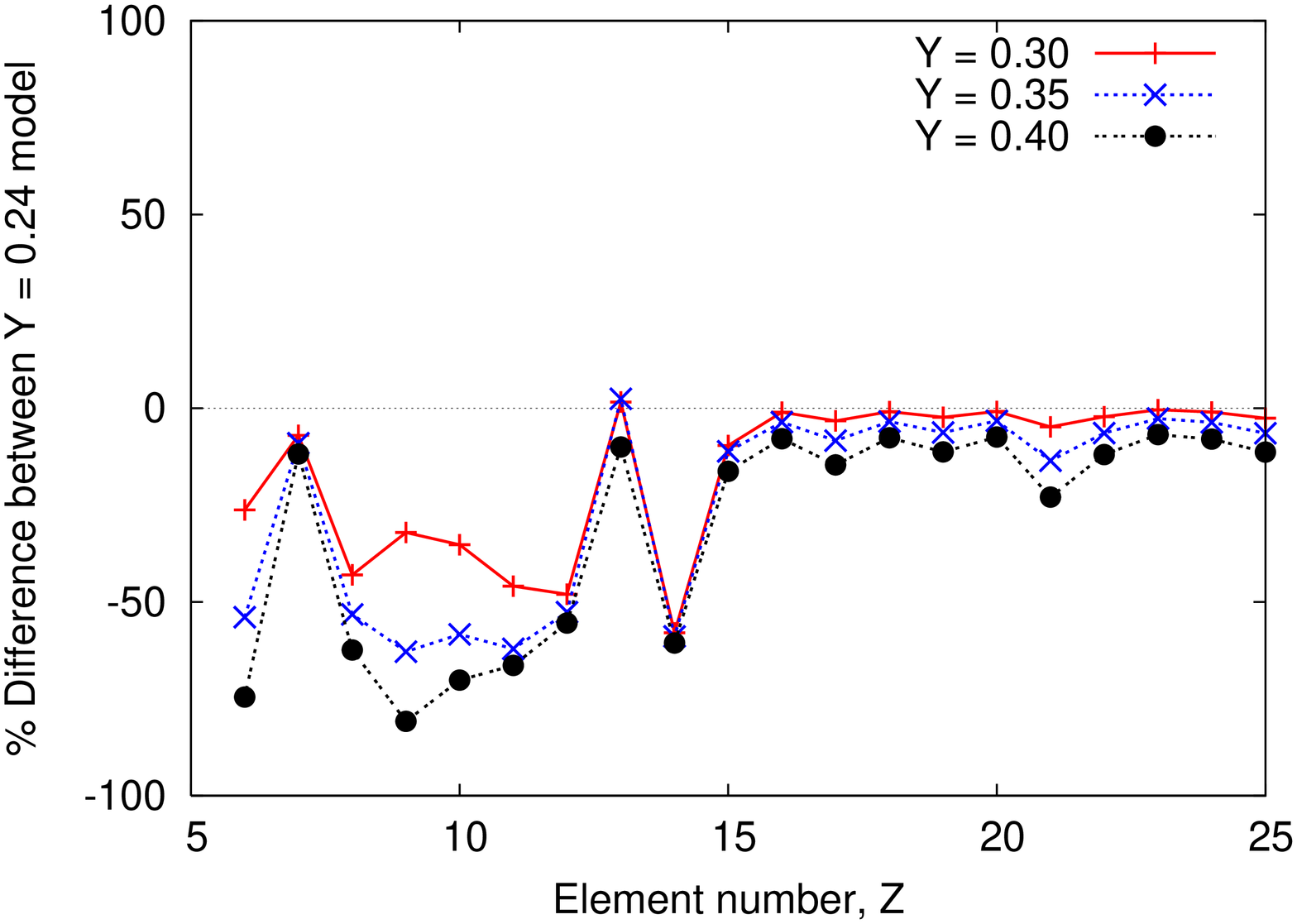}
\includegraphics[width=10cm,angle=270]{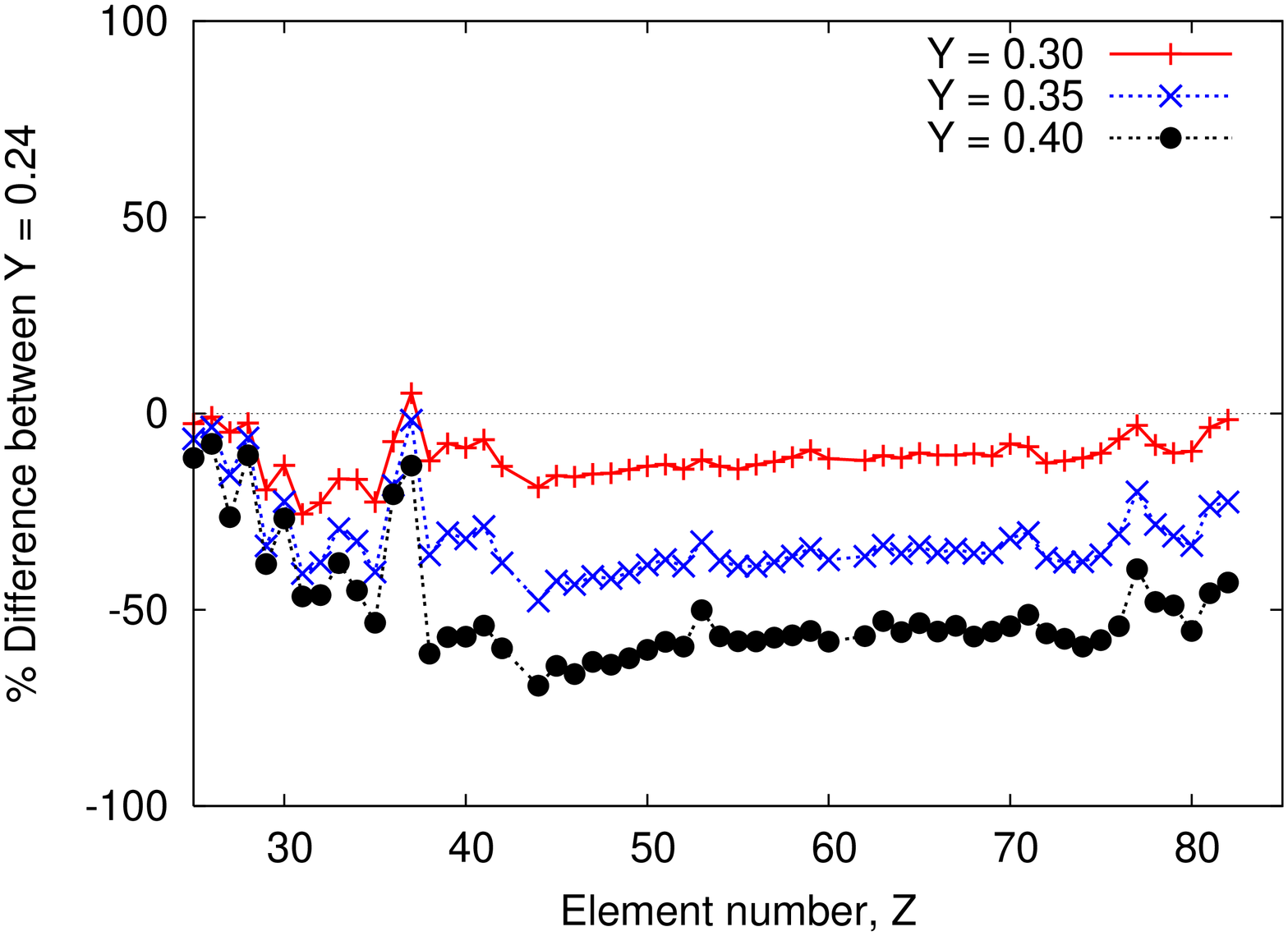}
\caption{Percentage difference between the mass of X expelled in the winds of the 1.7$\Msun$, $Z=0.0006$
models with varying $Y$ relative to the amount expelled in the primordial model 
with $Y = 0.24$. The upper panel shows the difference for elements lighter than Fe and the upper panel
for elements heavier than Fe. All models are calculated with the same $M_{\rm mix} = 0.002\Msun$.
\label{fig2}}
\end{figure}

\begin{figure}
\includegraphics[width=10cm,angle=270]{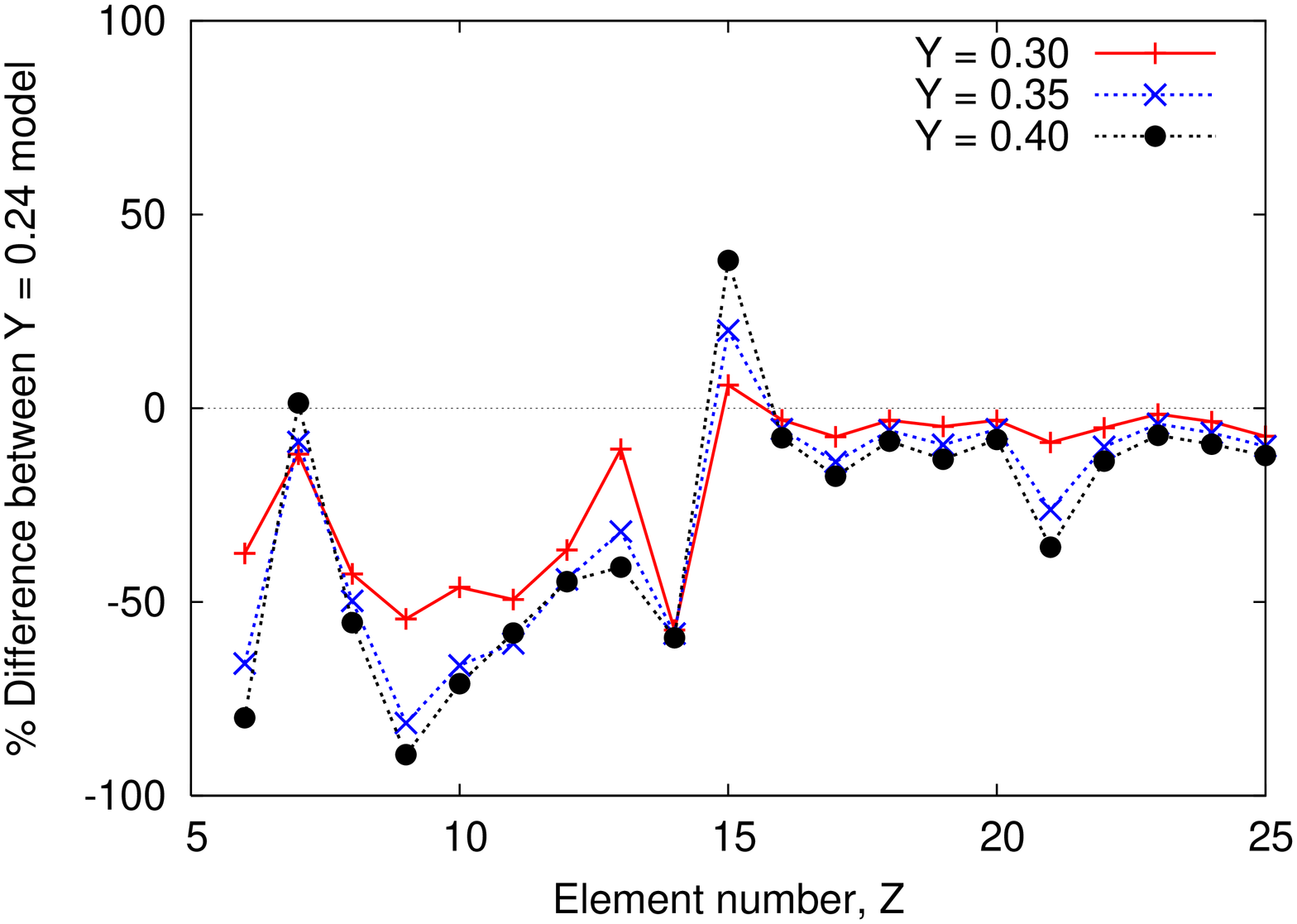}
\includegraphics[width=10cm,angle=270]{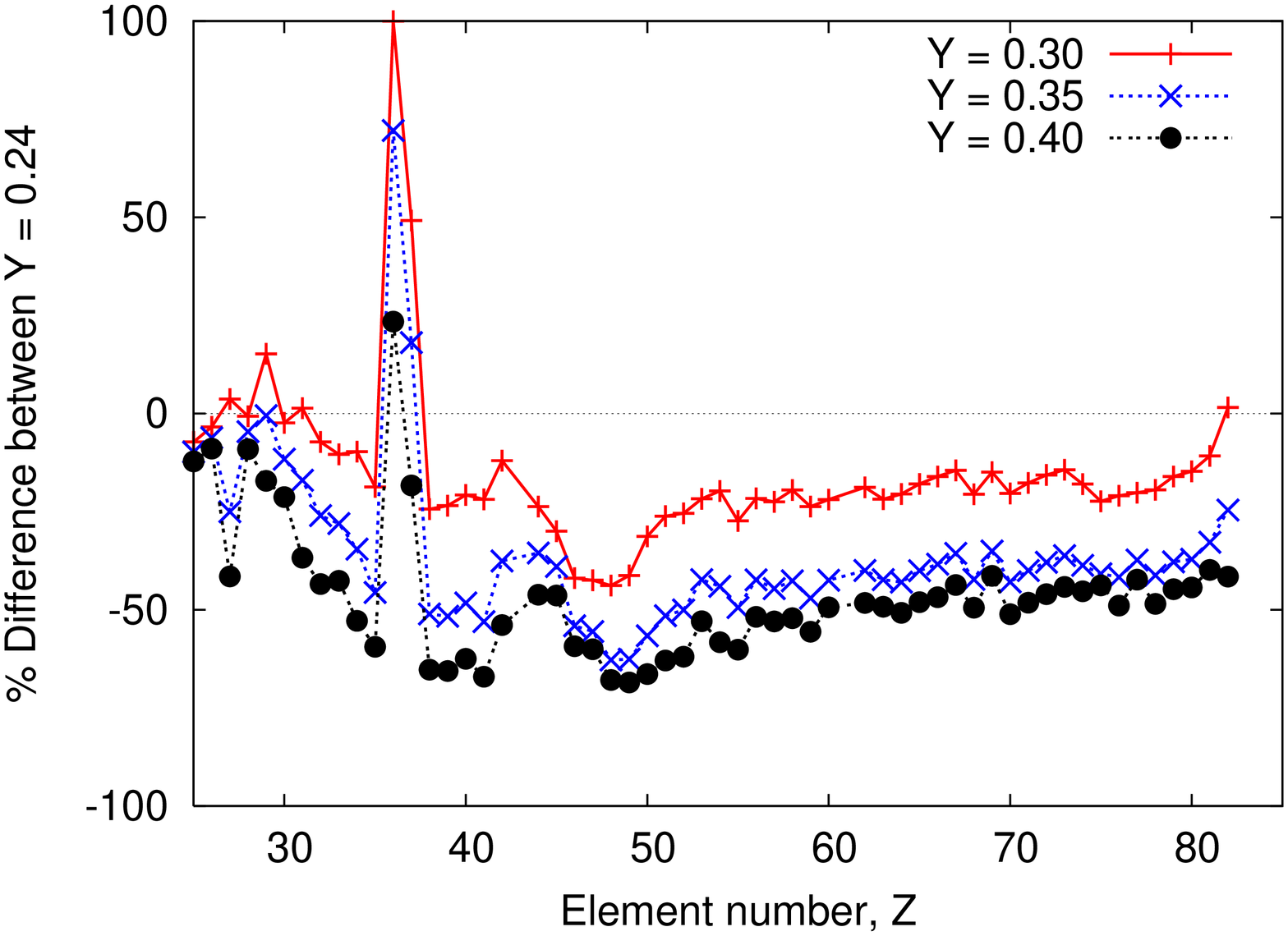}
\caption{Percentage difference between the mass of X expelled in the winds of the 2.36$\Msun$, $Z=0.0006$
models with varying $Y$ relative to the amount expelled in the primordial model 
with $Y = 0.24$. The upper panel shows the difference for elements lighter than Fe and the upper panel
for elements heavier than Fe. All models are calculated with the same $M_{\rm mix} = 0.002\Msun$.
\label{fig3}}
\end{figure}

In order to discuss how variations in helium affect the stellar yield calculations 
we show the percentage difference between the yields calculated from the helium-enriched models relative
to the yields from the $Y=0.24$ model for 1.7$\Msun$, $Z = 0.0006$ in Figure~\ref{fig2} and for
2.36$\Msun$, $Z = 0.0006$ in Figure~\ref{fig3}. 
Increasing the helium content by $\Delta Y = 0.1$ in low-mass AGB models 
results in a reduction  in the stellar yields by up to 65\% for C, 80\% for F, and roughly 
45\% for the $s$-process elements Ba and La.  While the yields of the first $s$-process 
peak elements Sr, Y, and Zr decrease by up to 50\%, the yields of Rb either do not change or increase.
 
Models with increasing helium do expel less material
into the ISM but it is only on the order of $\sim 10$\%. For example, the 1.7$\Msun$ model primordial
composition model expels 1.038$\Msun$ whereas the $Y=0.40$ model expels a total 0.937$\Msun$, a 
reduction of about 10\%. Figure~\ref{fig3} however shows that the yield differences are higher 
than 50\% between these two models for some elements. Elements most affected are those produced 
by AGB nucleosynthesis (e.g., C, F, $s$-process elements).  Intermediate-mass and Fe-peak elements 
are seen to vary at the level of $\approx 10$\%, as expected from variations in total 
expelled matter.

\begin{figure}
\includegraphics[width=10cm,angle=270]{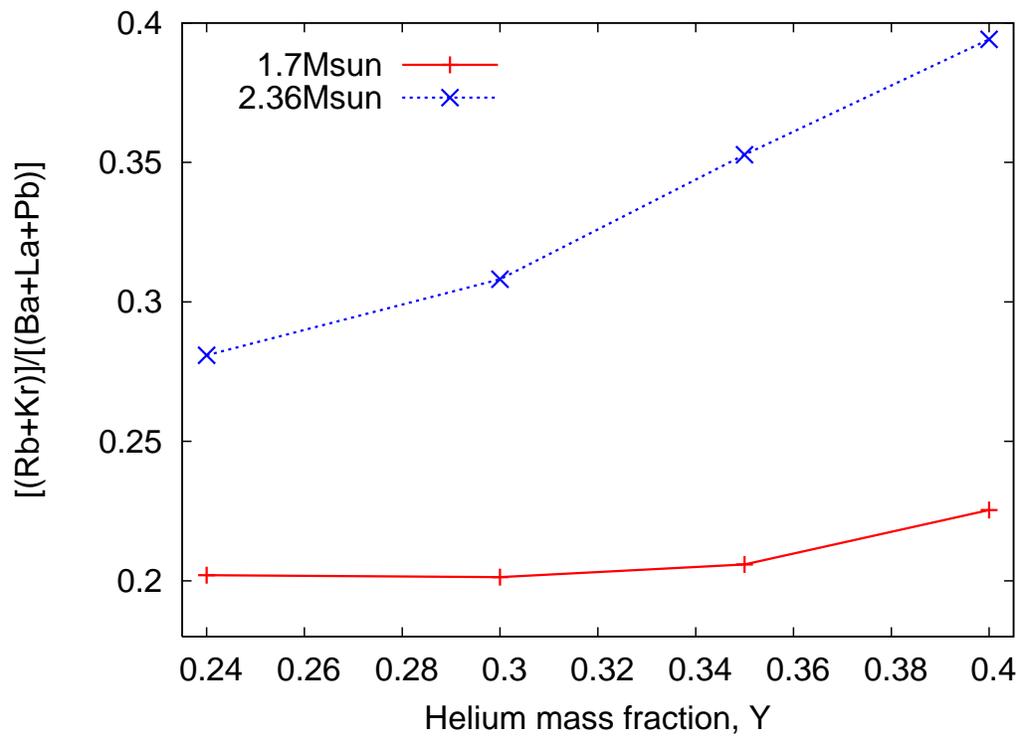}
\caption{The ratio of [(Rb$+$Kr)]/[(Ba$+$La$+$Pb)] from the 1.7$\Msun$ and 2.36$\Msun$
stellar models of $Z = 0.0006$ with an $M_{\rm mix} = 0.001\Msun$. Abundances are the average
in the ejected wind and calculated from the [X/Fe] ratios. 
\label{fig4}}
\end{figure}

Figures~\ref{fig2} and~\ref{fig3} show that the stellar yields of P, Cu, Kr, and Rb
show a reversed trend (i.e., an increase with increasing helium abundance). We further illustrate
this in Figure~\ref{fig4}, which shows the ratio of [(Rb$+$Kr)]/[(Ba$+$La$+$Pb)] from the 1.7$\Msun$
and 2.36$\Msun$ models with $M_{\rm mix} = 0.001\Msun$.  Out of these elements, 
Rb and Kr vary the most with yield increases of up to 50\%, with the Cu and P increases
being reasonably small (roughly 15\% and 40\% for Cu and P respectively, 
see Figures~\ref{fig2} and~\ref{fig3}).  Out of Kr and Rb, only Rb is observed 
in low-mass GC giants\footnote{Kr has been measured to be enhanced in the spectra of 
planetary nebulae \citep{sterling08}.}. 
The increase in the yields of P, Kr and Rb is much more pronounced for the 2.36$\Msun$
$Z = 0.0006$ case (Figures~\ref{fig3} and~\ref{fig4}) than in the 1.7$\Msun$ model. 
This is because helium enrichment causes higher He-shell temperatures which leads to 
partial activation of the \iso{22}Ne($\alpha,n$)\iso{25}Mg reaction 
during convective thermal pulses.

\subsection{Results for Heavy Elements}

\begin{figure}
\includegraphics[width=10cm,angle=270]{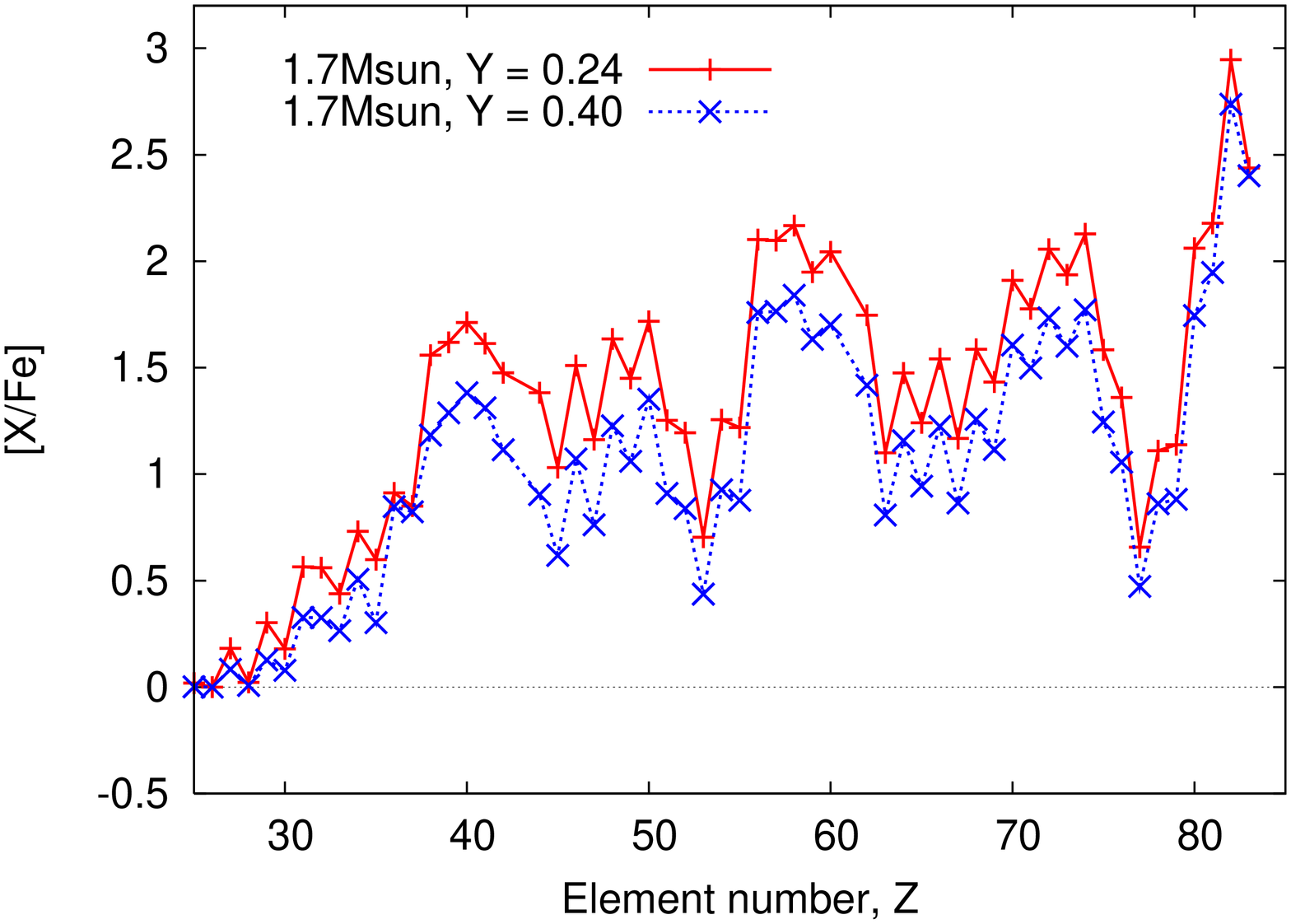}
\includegraphics[width=10cm,angle=270]{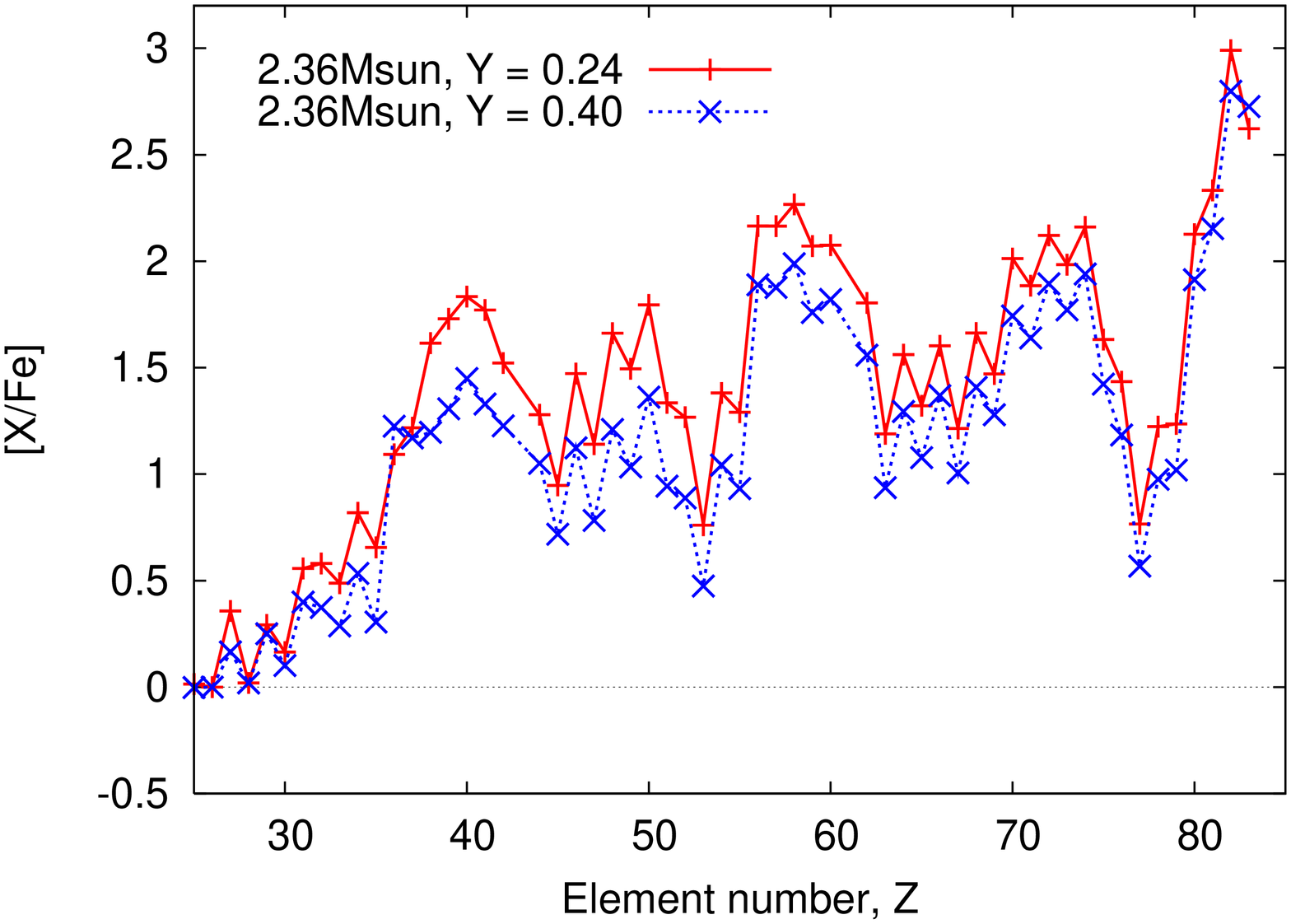}
\caption{Average predicted abundance in the ejected wind (in [X/Fe]) for elements heavier than iron from models of 
1.7$\Msun$, $Z = 0.0006$ (top panel) and 2.36$\Msun$, $Z = 0.0006$ (bottom panel). The 
1.7$\Msun$ and 2.36$\Msun$ models are calculated with the same $M_{\rm mix} = 0.002\Msun$
(see text for details).\label{fig5}}
\end{figure}

In Figure~\ref{fig5} we show the surface abundances in the ejected wind (in [X/Fe]) for elements heavier 
than iron for the 1.7$\Msun$ (top panel) and 2.36$\Msun$ (lower panel) models with a primordial helium 
composition and with $Y = 0.40$. The figure illustrates that models
with enriched helium display a typical $s$-process pattern associated with low-mass, 
low-metallicity AGB stars, albeit at lower [X/Fe] values than models with a primordial 
helium composition.  The lower [X/Fe] values show in Figure~\ref{fig5} are the result of less 
TDU material being mixed into the envelope during the TP-AGB lifetime.

We can measure if changing the initial helium content of the stellar evolutionary sequence
changes the $s$-process distribution. We can measure the shape of the distribution by comparing the
values of the $s$ process indicators: [Rb/Zr], [ls/Fe], [hs/Fe], [hs/ls], and [Pb/hs]. 
The ratios [Rb/Zr] and [ls/Fe] measure the level of enrichment at the first $s$-process peak around 
Y-Sr-Zr while the second measures the level enrichment at the second peak at around Ba.
The ratio  [hs/ls] is a particularly useful diagnostic because it is essentially independent of the
amount of TDU, mass-loss rate uncertainties and depends on thermodynamic conditions in the He-intershell,
which in turn determine the number of neutrons available for the $s$ process
\citep[e.g.,][]{busso01,bisterzo10,lugaro12}. 

We find that varying helium in the stellar evolutionary model changes the evolutionary behavior
and the structural details during the TP-AGB but it does not have a significant effect on the 
$s$-process distribution. For example, we obtain the same [hs/ls] $\approx 0.50$ in the 1.7$\Msun$, 
$Z = 0.0006$ models with $Y = 0.24$ and $Y = 0.40$ when using the same $M_{\rm mix} = 0.002\Msun$.
Similarly for [Pb/hs], which measures the 
amount of Pb synthesized relative to elements at the second peak around Ba, we obtain very similar 
values in the $Y=0.24$ and $Y=0.40$ models of [Pb/hs] = 0.82 and 0.95, respectively. 
These similar values indicate that the overall $s$-process distribution varies little between the 
1.7$\Msun$ models, regardless of their initial helium content and the total amount of TDU. 

This is also the case for the 2.36$\Msun$, $Z = 0.0006$ model although in the most helium-enriched
cases we see small increases of $\approx 0.1$ in the $s$-process indicator [hs/ls] and 
[Pb/hs] in models of the same partially mixed zone. These increases can be 
attributed to the the size of the partially mixed zone relative to the mass of the He-intershell. 
In the most helium enriched models the mass of the He-intershell decreases as shown in Table~\ref{table1}.
This means that the ratio of the mass of the resultant \iso{13}C pocket relative 
to the mass of the He-intershell is larger in the helium-enriched models. 

Models with higher helium content have higher He-shell
temperatures. For the 2.36$\Msun$ models the maximum temperature in the He-intershell is reached in the most
helium-enriched models with $Y \ge 0.35$ and many TPs have peak temperatures well above 300 MK. 
This means that there will be some activation of the \iso{22}Ne($\alpha,n$)\iso{25}Mg
neutron source and this increases the overall amount of $s$-process enrichments.  We can measure this by 
examining the [Rb/Zr] and [Rb/Sr] ratios. Low-mass AGB stars  produce more Sr and Zr than Rb as a 
result of the neutron density staying below $\lesssim 10^{8}$ n cm$^{-3}$ and the dominance of the 
\iso{13}C($\alpha,n$)\iso{16}O reaction \citep{busso01,abia02,vanraai12}. 
The [Rb/Zr,Sr] ratios are therefore an important observational diagnostic 
\citep[e.g.,][]{abia02,mcwilliam13}. Positive ratios indicate efficient operation of 
the \iso{22}Ne($\alpha$,n)\iso{25}Mg source whereas negative ratios indicate low-mass 
AGB stars and the operation of the \iso{13}C($\alpha$,n)\iso{16}O  reaction.

In the stellar models considered here, the [Rb/Sr] and [Rb/Zr] ratios are always negative.
The primordial 1.7$\Msun$ models give [Rb/Sr] $\approx -0.7$ and [Rb/Zr] $\approx -0.85$, 
where there is little variation with \iso{13}C pocket size. 
Focusing only on [Rb/Zr], an increase of $\Delta Y = 0.10$  increases the [Rb/Zr] ratio 
in all models except the 2.36$\Msun$ model with a small $M_{\rm mix} = 1\times 10^{-4}\Msun$. 
In the following we compare predictions from models with $Y=0.24$ and models with $Y=0.35$.
At 1.7$\Msun$, the difference in [Rb/Zr] between the helium-rich and primordial model is
0.09, 0.16, and 0.21, respectively for the three different choices for $M_{\rm mix}$ 
of 0.001, 0.002, and $0.004\Msun$. At 2.36$\Msun$, the [Rb/Zr] ratio first decreases by 
0.07 and then strongly increases by 0.38 and 0.36, respectively for $M_{\rm mix}$ 
equal to 0.0001, 0.001, and $0.002\Msun$.  

\begin{figure}
\includegraphics[width=10cm,angle=270]{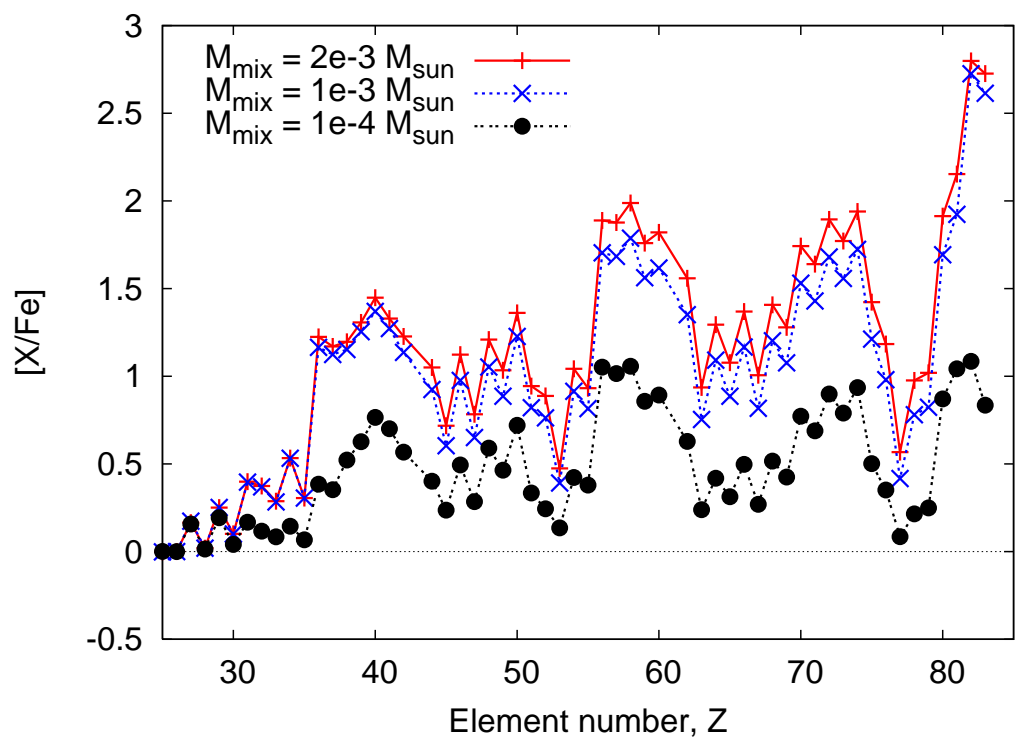}
\caption{Average predicted abundances in the ejected wind (in [X/Fe]) for elements heavier than iron from models of 
2.36$\Msun$, $Z = 0.0006$ with $Y=0.40$. Results are shown for the different sizes of $M_{\rm mix}$
= 0.002, 0.001, and 0.0001$\Msun$, respectively. \label{fig6}}
\end{figure}

The size of the partially mixed zone, $M_{\rm mix}$, determines the size of the \iso{13}C pocket that forms
in the He-intershell following a thermal pulse and consequently the number of neutrons available for the 
$s$ process.  In Figure~\ref{fig6} we show the abundances in the wind for the 2.36$\Msun$, $Z = 0.0006$ 
models of $Y=0.40$ with three different sizes for $M_{\rm mix}$.  We can draw two conclusions from 
Figure~\ref{fig6}. The first is that changing $M_{\rm mix}$ by a factor of $\approx 2$ leads to smaller 
changes to the overall level of predicted chemical enrichment relative to changing the helium abundance.
Second, large changes to $M_{\rm mix}$ of a factor of 10 or more result in the most striking changes to the
level of $s$ processing in the nucleosynthesis models. This is not entirely surprising as the $s$ process 
requires a \iso{13}C pocket for efficient activation in low-mass AGB stars 
\citep{gallino98,busso01,karakas07a}, regardless of the depth of TDU or other stellar evolutionary 
details that change when changing $Y$.

\section{Discussion and Conclusions} \label{sec:discussion}

Omega Centauri represents a suitable laboratory to test the predictions from helium-enhanced AGB yields. 
Observations reveal that this cluster has experienced a complex star-formation history with
likely many kind of polluters at work (supernovae, AGB of various masses, rotating massive stars, 
and/or binaries). 
The complexity of its chemical evolution, which occurred in a complex sequence of multiple and/or continuous
bursts of star formation, is reflected in many {\it peculiar} chemical patterns such as the large 
variation in metallicity, $s$-process elements, and helium, and O-Na/C-N anti-correlations
observed at various ranges of metallicities. The understanding of the whole observed chemical 
pattern in $\omega$~Cen is challenging, and requires complex models of chemical evolution 
that will eventually take into consideration not just the candidate polluters of various types,
but even different initial abundances for the same polluters. 

We apply the theoretical yields calculated here to $\omega$~Cen, focusing on the possible 
contribution to the chemical enrichment from low-mass AGB stars with both primordial and 
enhanced helium compositions.
The first generations of stars in $\omega$ Cen were low metallicity 
([Fe/H] $\lesssim -1.8$) with a primordial helium composition \citep[$Y \approx 0.24$;][]{piotto05}.
The AGB stars at various masses from this primordial stellar population
would have provided the ISM with large amounts of material converted into 
stars at various epochs. In particular, low-mass AGB stars from this stellar
generation would produce substantial amounts of $s$-process elements
(e.g., similar to the $Z$=0.0003 model yields presented here) on
relatively short times scales ($\sim$ 500 Myr if we consider stars of
$M \geq$ 2.3$\Msun$). Here we assume that subsequent generations of stars were formed out
of a higher metallicity gas that was also helium enriched. 
According to the models presented here, low-mass AGB stars from
these generations would expel roughly the same amount of material but
with yields that are much less enriched relative to their lower
metallicity and primordial helium counterparts. The stellar yields 
calculated here show a decrease of roughly 50\% for elements produced
by low-mass AGB stars when increasing helium by $\Delta Y = 0.10$.

What are the consequences? For elements produced by low-mass AGB stars such 
Ba and La the truncation in the stellar yields could explain the observed 
``ceiling'' in $s$-process elements seen by e.g., \citet{norris95}.
That is, the first generations expel a lot, especially relative to the initial 
amount of heavy elements present in the ISM. The second (and later) generations 
of helium-enriched stars expel much less Ba and La so there would only be a 
small incremental rise in the chemical evolution.  

If the chemical evolution of elements such as Ba, La and 
Pb are a consequence of their lower production in helium-enriched models, 
our models predict that the evolution of Kr and Rb are expected to rise, at least 
until the most helium-enriched AGB stars have formed.  
The sample by \citet{smith00} consists of Rb measurements for
seven stars covering the metal-rich range observed in $\omega$~Centauri, 
$-$1.5$<$[Fe/H]$< -$0.7, with three out the seven stars re-analysed from a 
previous study \citep{vanture94} to place the Rb abundances on a common scale.  
Despite the small number of stars and the limited range in metallicity, the
data display an increasing [Rb/Fe] with iron and unlike the behavior of the 
heavier $s$-process elements, the abundances of [Rb/Fe] in the metal-rich 
stars ([Fe/H]$\gtrsim -$1.0) seems to  keep increasing. The ratio of 
[Rb/Zr] shown in Fig.~14 from \citet{smith00} shows a possible 
increase with increasing [Fe/H] from $\approx -1.5$ to $-0.8$, consistent with our
models, although error bars are large and there are few stars 
\citep[see also Fig.~16 from][]{mcwilliam13}. Note that \citet{dorazi11} also measure 
Rb but only for one star. They found [Rb/Zr] $= -0.65$, consistent with 
low-mass AGB nucleosynthesis.

If this chemical pattern for Rb and Zr could be confirmed for a 
statistically significant sample of stars, it may qualitatively
support the idea that helium-enhanced low-mass AGB stars have contributed to 
the chemical enrichment of $\omega$~Cen. 
The helium-enriched AGB stars would also evolve much more
quickly than their primordial counterparts and would have time to
expel Rb (and less Sr, Zr etc) into the ISM before the star formation of $\omega$~Cen
has finished. 

As previously mentioned, the observed chemical pattern for the light elements 
C, N, O, and Na is very complex in $\omega$~Cen, as these elements show 
large variations at all metallicities \citep[see Fig.~7 in][]{marino12a}.
A significant contribution to the evolution of these elements may be
a result of the high-temperature H-burning that occurred in intermediate mass 
AGB stars as outlined by \citet{dantona11}.
The helium-enhanced lower mass models are expected to significantly 
under-produce these elements (e.g., C) relative to primordial helium models. Regardless,
it is difficult to qualitatively compare our predictions with observations, 
as the abundance of these elements is likely the result of a mixture of 
polluters that may contribute in opposite directions to the ISM enrichment. 
We point out here that the abundance of [C/Fe] is not observed to
decline with metallicity. The [C/Fe] increase with [Fe/H] is less pronounced 
for Na-rich (likely He-rich) stars \citep{marino11}, and this could
qualitatively support the idea that low-mass, helium-rich AGB yields
produce less carbon.

The total CNO over Fe predicted from AGB with different helium
enhancement varies by $\sim$0.6~dex, with lower values for highly
helium-enhanced AGB models. A C$+$N$+$O increase with metallicity has been
observed in $\omega$~Cen \citep{marino12a}, but it is considerably
lower (by $\sim$1~dex) than what is predicted from the helium-normal, low-mass AGB
models. Unaccounted for effects such as dilution may also decrease the 
CNO abundance with respect to the predicted yields. We note that 
the increase of C$+$N$+$O is expected to drop when applying 
helium-enhanced AGB yields.

Knowledge of the timescales over which star formation occurred 
in $\omega$~Cen is fundamental to understanding the chemical evolution of this cluster.
The heavy to light $s$-process elemental ratio, [hs/ls], is sensitive 
to the progenitor AGB masses \citep[e.g.,][]{busso99}, but chemical 
abundances reported in the literature do not provide unique information 
regarding the AGB mass range that may have contributed to the chemical enrichment 
of this cluster. \citet{smith00} found that the increase in heavy
$s$-process elements, such as Ba, La, and Nd is larger than that in the 
lighter elements Y, Zr, and Mo, suggesting that the observations
are better reproduced by lower mass AGB polluters of $\approx 1.5\Msun$. 
Consequently, the star formation for $\omega$~Cen would have proceeded
for a long period of at least $\approx $1--3~Gyrs. On the other hand,
\citet{dorazi11} found that light $s$-process elements Y and Zr
vary more than the second-peak ones La and Ce, suggesting that the
main $s$-process component active in $\omega$~Cen tends towards masses
$\approx 3\Msun$, reducing enrichment timescales from 1-3~Gyr, to a few 
hundred million years. 

From the lifetimes given in Table~\ref{table1}, the helium-enriched stellar models 
of $Z=0.0006$ follow the relationship,
\begin{equation}
\tau_{\rm stellar} \appropto (M)^{-2.69} {\times} \exp{\left[-5.43(Y-0.24)\right]},
\end{equation}
This means that either an increase in $Y$ of 0.05 or of the initial 
mass by 11\% will decrease the lifetime of the star by $\sim$24\%.

Following the \citet{dercole10} scenario for the mono-metallic GC NGC\,2808, 
extremely helium-enhanced stars are expected to form immediately after the primordial stellar
generation, and, later on, mildly helium-enhanced stars form from diluted
material. If a similar scenario is applicable to $\omega$~Cen, we expect the
contribution from higher mass AGB that produced {\it extremely} He-enhanced
stars to have occurred early in the cluster chemical enrichment history.  Later on, 
lower-mass helium-enhanced AGB stars would have formed, evolving much faster than 
their lower-helium counterparts.

Our results imply that the contribution from these polluters is expected to occur
even if the evolution of the cluster had been confined to a few hundred Myrs and 
up to $\lesssim$2~Gyrs. This would diminish the apparent discrepancy between recent 
scenarios for a fast cluster evolution \citep[e.g.,][]{dantona11}, supported by 
recent isochrone fitting taking into account helium and CNO variations 
\citep[e.g.,][]{herwig12,joo13}, and the slow enrichment expected from the chemical
pattern of $s$-process elements. 

One caveat to all of this is that the decrease in the lifetime means that the
helium-enriched AGB stars might actually be contributing more gas. 
For example, if we assume that both stellar populations have a Salpeter 
initial mass function; that there was 100 million years between the first and second 
generations and that the second generation absorbed all the gas; that the last supernovae 
in the second generation took 40 Myr to occur; that $\Delta Y = 0.10$; that the first
generation and second generation have equal mass once the last supernovae from the 
second generation has gone off; then we get that the fraction of gas that is from 
second generation stars $\sim 1$ billion years into the cluster's life is 50\%. 
So in other words, it is plausible that half the non-pristine gas from the 
third generation comes from the second generation.  That would lead to a lower 
required amount of dilution by pristine gas in chemical evolution models of 
globular clusters with three generations. However, note that the total dilution 
required might still increase. Even though the yields of second generation stars 
are lower, these stars will start off with enriched abundances of some elements, 
and so the final released abundance might be higher.

Large helium enhancements have also been inferred for other Galactic globular
clusters besides $\omega$ Cen including M22 \citep{joo13}, NGC 2808 
\citep{dantona05,pasquini11,marino14}, NGC 2419 \citep{dicris11}, Terzan 5 
\citep{dantona10},  NGC 6388 and NGC 6441 \citep{gbusso07,caloi07,yoon08}. 
With the exception of M22, none of these clusters have been demonstrated to show 
clear evidence for star-to-star variations in $s$-process elements which suggests that 
low-mass AGB stars have not had time to contribute to the chemical evolution of these 
systems. The effect of helium enrichment on the 
stellar yields of intermediate-mass AGB stars is unknown but if it has a similar effect as
in low-mass AGB models, the yields are likely to be strongly reduced. Helium enrichment
could help increase the mass-loss rate via an increase in the stellar luminosity and reduce 
the level of chemical enrichment from stellar models. This may also help truncate the AGB 
lifetime without the need for faster mass-loss rates \citep[e.g.,][]{dorazi13a}.  

Stars in the cluster NGC 1851 show a spread in the abundances of C$+$N and $s$-process 
elements \citep{yong09,lardo12} but helium variations in NGC 1851 are likely smaller 
than $\Delta Y=0.10$ \citep[where $\Delta Y \lesssim 0.06$,][]{joo13}.  M22 shows a spread
in iron and $s$-process elements \citep{dacosta09,marino09,roederer11}, and \citet{joo13} determine
that $\Delta Y = 0.09$ from a comparison between the CMD and theoretical stellar models.
While the age spreads of M22 and NGC 1851 are smaller than $\omega$ Cen, at $\approx 300$~Myr
for M22 \citep{marino12b,joo13} and 300--500~Myr for NGC 1851 \citep{cassisi08,milone08,joo13},
the abundance spreads suggest the contribution from low-mass AGB stars, although perhaps 
higher in mass ($\sim 3\Msun$) than we considered here.  For these clusters, stellar yields 
similar to ours that take into account variations of helium may be necessary in order to 
reproduce the chemical evolution of these systems.

Globular clusters are not the only systems with claims of extremely helium-enhanced 
subpopulations. An increase in helium for the Galactic bulge has also been suggested
based on observations of the red giant branch bump \citep{nataf11}, and the discrepancy between 
its photometric and spectroscopic turnoff ages \citep{nataf12}. Helium-enhanced subpopulations 
have also been suggested for ellipticals and galactic spheroids \citep{chung11,bekki12} 
as a solution to the phenomenon of the UV-upturn \citep{code79,brown04,atlee09}.
Given that in both the case of the Galactic bulge and of extragalactic spheroids, 
the helium enrichment is associated with metal-rich populations, it may be the case that 
further insights could be gleaned from fitting abundance trends predicted by chemical 
evolution models to measurements made of metal-rich stars and of the integrated light 
of metal-rich systems. Further investigations of globular clusters could be used 
to gauge the reliability of helium-enhanced AGB models, and thus their potential 
applicability to field populations. 

Finally, our results confirm that a new dimension needs to be included when considering
stellar yields from stars and the chemical evolution of galaxies and stellar systems.
It may not be enough to simply evolve grids of stellar evolutionary sequences covering a range
in mass and metallicity. We show  that variations in helium mass fraction have a significant impact
on the stellar yields and may be an important third parameter. 

\acknowledgments

The authors would like to thank the referee for providing a comprehensive and
highly constructive report. 
DN would like to thank Paolo Ventura for useful discussions on this topic.
AIK thanks Maria Lugaro for use of the nuclear network used in this study and is grateful 
for the support of the NCI National  Facility at the ANU. AIK was supported through an
Australian Research Council Future Fellowship (FT110100475).

\clearpage

\clearpage

\clearpage

\clearpage

\clearpage

\clearpage

\begin{table}
 \begin{center}
  \caption{Details of the stellar evolutionary sequences.
 \label{table1}}
  \vspace{1mm}
   \begin{tabular}{l|lllll}
\tableline\tableline
\multicolumn{6}{c}{1.7$\Msun$ models} \\ \tableline
     & $Y=0.24$   & $Y = 0.24$ &  $Y = 0.30$ & $Y = 0.35$ & $Y=0.40$ \\
     & $Z=0.0003$ & $Z=0.0006$ &  $Z=0.0006$ & $Z=0.0006$ & $Z=0.0006$ \\
\tableline
$\tau_{\rm ms}$ & 871 & 880 & 629 & 481 & 367 \\
$\tau_{\rm rgb}$ & 252 & 276 & 187 & 128 & 81.5 \\
$\tau_{\rm core He}$ & 124 & 110 & 118 & 114 & 107 \\
$M_{\rm c}^{\rm 1st}$ & 0.574 & 0.568 & 0.607 & 0.655 & 0.726 \\
$M_{\rm c}^{\rm TDU}$ & 0.587 & 0.586 & 0.627 & 0.672 & 0.741 \\
Number of TPs & 16 & 17 & 16 &  15 & 13 \\
$M_{\rm dredge}^{\rm tot}$ & 0.082 & 0.071 & 0.0495 & 0.030 & 0.016 \\
$\lambda_{\rm max}$ & 0.64 & 0.62 & 0.60 & 0.58 & 0.57 \\
Maximum $T_{\rm TP}$ & 287 & 285 & 288 & 289 & 292 \\
Maximum $M_{\rm Heshell}^{\rm f}$ & 0.0186 & 0.0160 & 0.0150 & 0.0110 & 0.0074 \\
Final stellar mass & 0.665 & 0.662 & 0.692 & 0.806 & 0.980 \\ 
Final core mass &  0.665 & 0.662 & 0.683 & 0.712 & 0.763 \\
Final $M_{\rm env}$ & 0.0 &  0.0 & 0.01 & 0.09 & 0.22 \\
$\tau_{\rm stellar}$ & 1418 & 1436 & 943.3 & 730.7 & 563.2 \\ 
\tableline
\multicolumn{6}{c}{2.36$\Msun$ models} \\ \tableline
     & $Y=0.24$   & $Y = 0.24$ &  $Y = 0.30$ & $Y = 0.35$ & $Y=0.40$ \\
     & $Z=0.0003$ & $Z=0.0006$ &  $Z=0.0006$ & $Z=0.0006$ & $Z=0.0006$ \\
\tableline
$\tau_{\rm ms}$ & 348 & 373 & 280 & 218 & 169 \\
$\tau_{\rm rgb}$ & 46.7 & 40.0 & 27.2 & 20.6 & 15.8 \\
$\tau_{\rm core He}$ & 99.6 & 117 & 88.1 & 68.2 & 52.5 \\
$M_{\rm c}^{\rm 1st}$ & 0.682 & 0.658 & 0.733 & 0.803 & 0.864 \\
$M_{\rm c}^{\rm TDU}$ & 0.688 & 0.667 & 0.736 & 0.808 & 0.868 \\
Number of TPs & 19 & 18 & 16 & 17 & 20 \\
$M_{\rm dredge}^{\rm tot}$ & 0.159 & 0.158 & 0.090 & 0.045 & 0.027\\
$\lambda_{\rm max}$ & 0.94 & 0.92 & 0.95 & 0.94 & 0.90 \\
Maximum $T_{\rm TP}$ & 319 & 318 & 319 & 329 & 327 \\
Maximum $M_{\rm Heshell}^{\rm f}$ & 0.0138 & 0.0161 & 0.0115 & 0.0063 & 0.0039 \\
Maximum $T_{\rm bce}$ & 10.7 & 8.05 & 14.7 & 23.9 & 40.0 \\
Final stellar mass & 0.832 & 0.816 & 1.074 & 1.274 & 1.493 \\
Final core mass &  0.708 & 0.693 & 0.752 & 0.818 & 0.879 \\
Final $M_{\rm env}$ & 0.123 & 0.123 & 0.32 & 0.46 & 0.61 \\
$\tau_{\rm stellar}$ & 506.6 & 541.3 & 402.6 & 312.4 & 241.1 \\
\tableline \tableline
 \end{tabular} 
\tablecomments{Stellar lifetimes are in $\times 10^{6}$ years (Myr); masses in solar masses ($\Msun$)
and temperatures in $\times 10^{6}$K (MK).}
 \end{center}
\end{table}

\clearpage

\begin{deluxetable}{ccrrrrr}
\tabletypesize{\scriptsize}
\tablecaption{An example of the on-line data tables $Z = 0.0006$ models. We show the
yields for the 1.7$\Msun$ model with $Y=0.24$ and $M_{\rm mix} = 0.001\Msun$.\label{table2}}
\tablewidth{0pt}
\tablehead{
\colhead{\# El} & \colhead{$Z$} & \colhead{$\log$ e(X)} & \colhead{[X/H]} & 
\colhead{[X/Fe]} & \colhead{X(i)} & \colhead{Mass(i)}
}
\startdata
  p & 1 & 12.000000 & 0.000000 & 0.000000 & 7.10959E-01 & 7.37975E-01 \\
 he & 2 & 11.024580 & 0.094580 & 1.470952 & 2.98779E-01 & 3.10132E-01 \\
  c & 6 & 9.187603  & 0.717603 & 2.093975 & 1.30487E-02 & 1.35445E-02  \\
  n & 7 & 7.051185 & $-$0.818815 & 0.557558 & 1.11151E-04  & 1.15375E-04 \\
  o &  8 & 7.919069 & $-$0.810930 & 0.565442 & 9.36582E-04 & 9.72173E-04 \\
  f  & 9 &  5.169071 & 0.749071 & 2.125443 & 1.97771E-06 & 2.05287E-06 \\
 ne & 10 &  7.795207 & $-$0.174793 & 1.201580 & 8.88203E-04 & 9.21955E-04 \\
 na & 11 &  5.719754 & $-$0.520246 & 0.856126 & 8.50504E-06 & 8.82823E-06 \\
\sidehead{...}
\sidehead{\#[Rb/Zr], [ls/Fe], [hs/Fe], [hs/ls], [Pb/hs]\tablenotemark{a}}
\sidehead{$-$0.8824   1.4041   1.8718   0.4678   0.7845}
\enddata
\tablenotetext{a}{We also show the last two lines of the yield table for this particular model.}
\tablecomments{Each table starts with a header
that lists the initial mass, metallicity ($Z$), helium content ($Y$), size of 
$M_{\rm mix}$, the final core mass, and the total amount of 
mass ejected.  All masses are in solar units.}
\end{deluxetable}

\begin{table}
\begin{center}
\caption{Representative abundance ratios calculated from the average composition
in the wind for a selection of the $Z=0.0006$ stellar models with different $Y$.
\label{table3}}
\begin{tabular}{lcccccccccc}
\tableline\tableline
Mass & $Y$ & [C/Fe] & [N/Fe] & [F/Fe] & [Na/Fe] & [Kr/Fe] & [Rb/Fe] & [ls/Fe] &
[hs/Fe] & [Pb/Fe] \\
\tableline
1.70 & 0.24 & 2.09 & 0.56 & 2.13 & 0.86 & 0.68 & 0.61 & 1.41 & 1.87 & 2.66 \\
1.70 & 0.40 & 1.55 & 0.53 & 1.41 & 0.39 & 0.64 & 0.64 & 1.19 & 1.60 & 2.53 \\
2.36 & 0.24 & 2.21 & 0.52 & 2.49 & 1.10 & 0.86 & 0.99 & 1.50 & 1.95 & 2.73 \\
2.36 & 0.40 & 1.58 & 0.54 & 1.51 & 0.64 & 1.17 & 1.12 & 1.26 & 1.72 & 2.72 \\
\tableline
\end{tabular}
\tablecomments{Abundance predictions are shown for models with the same value 
of $M_{\rm mix} = 0.001\Msun$.}
\end{center}
\end{table}

\clearpage

\bibliographystyle{apj}
\bibliography{apj-jour,library}


\end{document}